\documentclass{elsart}
\usepackage{bm,epsfig,amssymb,amsmath,graphicx}

\newcommand{\bea}{\begin{eqnarray}}
\newcommand{\eea}{\end{eqnarray}}
\def\lag{\langle}
\def\rag{\rangle}

\def\be{\begin{equation}}
\def\ee{\end{equation}}

\begin{document}

\begin{frontmatter} 
\begin{minipage}[t]{16cm}{
\begin{center}
\title{$T-\mu$ phase diagram of the chiral quark model 
from a large flavor number expansion}
\author[BME]{A. Jakov\'ac},
\ead{jakovac@planck.phy.bme.hu}
\author[ELTE,now1]{A. Patk{\'o}s},
\thanks[now1]{Department of Atomic Physics}
\ead{patkos@ludens.elte.hu}
\author[HAS1]{Zs. Sz{\'e}p},
\ead{szepzs@antonius.elte.hu}
\author[ELTE,HAS2,now2]{P. Sz{\'e}pfalusy}
\thanks[now2]{Department of Physics of Complex Systems}
\ead{psz@galahad.elte.hu}

\address[BME]{
Hungarian Academy of Sciences and Budapest University of Technology
and Economics, Research Group ``Theory of Condensed Matter'',
H-1521 Budapest, Hungary}
\address[ELTE]{E{\"o}tv{\"o}s University, H-1117 Budapest, Hungary}
\address[HAS1]{Research Group for Statistical Physics of the
Hungarian Academy of Sciences, \\H-1117 Budapest, Hungary}
\address[HAS2]{Research Institute for Solid State Physics and Optics,
Hungarian Academy of Sciences, H-1525 Budapest, Hungary}

\begin{abstract}
The chiral phase boundary of strong matter is determined in the $T-\mu$ plane 
from the chiral quark model, applying a non-perturbatively renormalised 
treatment, involving chains of pion-bubbles and 1-loop fermion contributions. 
In the absence of explicit symmetry breaking the second order portion of the 
phase boundary and the location of the tricritical point (TCP) are determined 
analytically. Sensitivity of the results to the renormalisation scale
is carefully investigated. The
softening of the $\sigma$-pole near the second order transitions is confirmed.
\end{abstract}

\begin{keyword}
large N approximation, tricritical point, finite density
\PACS 11.10.Wx \sep 11.30.Rd \sep 25.75.Nq 
\end{keyword}

\end{center}
}
\end{minipage}
\end{frontmatter}

\section{Introduction}
The chiral phase structure of QCD at nonzero quark density receives renewed 
interest with the advent of new lattice techniques 
\cite{fodor02a,fodor02,karsch02,philipsen02,lombardo03,katz03}. 
Recent investigations concentrate on the transformation of the temperature 
driven crossover into first order transitions with increasing bario-chemical 
potential. In the present letter we shall address the same issue.

The lattice approach is an exact tool, but requires thorough investigation of 
finite quark mass and lattice spacing effects. In particular, the chiral limit 
with zero mass pions is beyond the access of present numerical QCD techniques 
(see, however \cite{wiese03}). An interesting attempt for a 
``phenomenological'' interpretation of lattice data in the hadronic phase was 
recently proposed. A hadronic resonance model with bag-motivated quark mass
dependence of the hadron spectra describes fairly well the thermodynamics of 
lattice QCD with unnaturally heavy pseudoscalar mesons in the low temperature 
phase \cite{redlich03a,redlich03b} (see also \cite{dumitru03}).

Extensive literature exists on the phase structure of effective theories 
matched to phenomenological data. Following the pioneering work of Barducci 
{\it et al.} \cite{barducci94} detailed studies were published in 
Nambu-Jona-Lasinio models
\cite{berges99,scavenius01,fujii03} and in the linear $\sigma$ model
\cite{hatsuda99,lenaghan00,scavenius01}.
  
These investigations do not go beyond some version of the quasi-particle 
approach, e.g. the thermodynamical potential is approximated by a sum of free 
particle contributions with masses obtained from some sort of selfconsistent 
gap equations. The renormalisation of the selfconsistent set of equations of 
state (EoS) and propagators represents a notoriously difficult problem. 
In particular in the broken phase the UV divergences of the selfenergies have 
temperature and $\mu$-dependent coefficients, also depending on the chiral 
condensate. In general, subtraction of such terms defines oversubtracted
renormalisation schemes. In higher orders of the perturbative solution one 
expects that resummed or improved expansions demonstrate the cancellation 
of the  apparent temperature/$\mu$/condensate-dependent counterterms
\cite{collins87}. Despite promising recent progress \cite{blaizot03,vanhees02} 
no general enough implementation is available to date for obtaining a 
renormalised set of equations in which resummed propagators could be used. 
Therefore it is hard to assess the accuracy of the predictions obtained from 
quasi-particle descriptions.

In the present letter we extend our recent treatment of the pure $O(N=4)$
model in the approximation of large number of Goldstone bosons ($N$)
\cite{szep02} to the chiral quark model at finite $\mu$. The effect of
the constituent quarks will be represented in this letter by 1-loop 
contributions. We shall find fermion effects which are of 
${\mathcal O}(1/\sqrt{N})$. A resummed renormalisation scheme will be
proposed for the coupled set of EoS, and the $\pi$ and $\sigma$
propagators. Finally, the renormalised equations will be analysed in
the chiral limit, with Goldstone's theorem being consistently obeyed.
Since $N=4$ is only moderately large, the fermionic contributions will
not be considered in the numerical solution as perturbations of the
leading large $N$ solution. They rather will be included fully 
into the equations, and we shall search for a new (non-iterative) solution. 
The phase boundary will be fully traced in the $T-\mu$ plane.

The parameters of the model will be fixed with reference to the 
$T=0$ mass and width of the $\sigma$ particle \cite{szepfalusy02}. 
It turns out that the renormalisation scale influences very
  strongly the shape of the spectral function in the sigma channel. The
  reason for this phenomenon is found in the appearance of extra
  unphysical poles on the physical Riemann sheet in a certain range of the
renormalisation scale. After restricting the investigation to the 
physically allowed region, we decided to scan through a large
domain of the renormalisation scale with the aim to see the sensitivity
of the phase boundary in the $T-\mu$-plane to its choice. 
We find that $T_c(\mu=0)$ is fairly close
 to the range of critical temperatures measured in two-flavor QCD with
different fermion discretisations.
The coordinates of TCP are in good qualitative agreement with its earlier 
determinations from effective models \cite{scavenius01}, 
implying $T_{TCP}\leq 70$ MeV independently of the renormalisation scale.

\section{The model and its equation of state}
In the definition of the model we display explicitly the flavor
factors:
\be
L[\sigma,\pi^a,\psi]=L_M[\sigma,\pi^a]+L_F[\sigma,\pi^a,\psi]+
\delta L_{ct}[\sigma,\pi^a,\psi],
\ee
where
\bea
L_M&=&-\left[\frac{\lambda}{24}\Phi^4+\frac{1}{2}m^2\Phi^2\right]N-
\left[\frac{\lambda}{6}\Phi^3+m^2\Phi\right]\sigma\sqrt{N}
\nonumber\\
&+&\frac{1}{2}\bigl[\partial^\mu\sigma (x)\partial_\mu \sigma (x)+\partial^\mu
\pi^a(x)\partial_\mu\pi^a(x)\bigr]-
\frac{1}{2}m^2\bigl[\sigma^2(x)+\pi^a(x)\pi^a(x)\bigr]\nonumber\\
&-&\frac{\lambda}{12}\Phi^2\bigl[3\sigma^2(x)+\pi^a(x)\pi^a(x)\bigr]
\nonumber\\
&-&\frac{\lambda}{6\sqrt{N}}\Phi\bigl[\sigma^3(x)+\sigma
(x)\pi^a(x)\pi^a(x)\bigr]-
\frac{\lambda}{24N}\bigl[\sigma^2(x)+\pi^a(x)\pi^a(x)\bigr]^2.\nonumber\\
L_F&=&\bar \psi(x)\left[i\partial^\mu\gamma_\mu-m_q-\frac{g}{\sqrt{N}}\left(
\sigma (x)+i\sqrt{2N_f}\gamma_5T^a\pi^a(x)\right)\right]\psi(x),
\eea
$\delta L_{ct}[\sigma,\pi^a,\psi]$ is the countertem Lagrangian.
In the large $N$ limit one has $N_f=\sqrt{N}$. The above form refers
explicitly to the broken symmetry phase with the condensate giving
rise also to the mass of the constituent fermion, which stays finite
in the large $N$ limit:
\be
\sigma (x)\rightarrow \sqrt{N}\Phi+\sigma (x),\qquad
m_q=g\Phi.
\ee
It is noteworthy that the linear $\sigma$ model has a second independent 
quartic coupling for $N_f> 2$, which we fix for simplicity at zero, 
since we wish to evaluate the expressions derived in the large $N$ 
approximation for $N=4$. Also, in this letter we work in the chiral limit, 
that is no explicit symmetry breaking term is introduced into the Lagrangian. 
The chemical potential for fermions is introduced into the fermionic piece of 
the Lagrangian by the replacement: $\partial_t\rightarrow \partial_t-i\mu$.

The form of EoS to ${\mathcal O}(1/\sqrt{N})$ with $N_c$ colored quarks 
is the following:
\be
\sqrt{N}\Phi\left[m^2+\frac{\lambda}{6}\Phi^2+
\frac{\lambda}{6N}\lag\pi^a\pi^a\rag_M\right]
+\frac{gN_c}{\sqrt{N}}\lag\bar\psi \psi\rag=0.
\label{Eq:EoS}
\ee
It is obvious that the fermionic contribution is ${\mathcal O} (1/\sqrt{N})$. 
The fermion tadpole will be evaluated using the tree-level mass $m_q$. 
The subscript $M$ at the expectation value of the composite operator 
$\vec\pi^2(x)$ refers to an effective pion tadpole computed with propagator 
mass $M$, which sums the contribution of all super-daisy type diagrams 
\cite{dolan74}. The corresponding pion propagator is determined from the 
Schwinger-Dyson equation as $iG_\pi^{-1}(p^2=M^2)=0$: 
\be
iG_\pi^{-1}(p)=p^2-m^2-\frac{\lambda}{6}\Phi^2
-\frac{\lambda}{6N}\lag\pi^a\pi^a\rag_M-B_{1\psi}(p).
\label{Eq:pion}
\ee
Here $B_{1\psi}$ is the contribution of the fermionic bubble to the
self-energy of the pion. By an algebraic transformation of its integrand
it can be related to the fermionic tadpole as follows:
\be
B_{1\psi}(p)=\frac{gN_c}{N\Phi}\lag\bar\psi \psi\rag+\frac{2g^2N_c}
{\sqrt{N}}p^2 I_F(p,m_q),
\label{Eq:bubble-tadpole}
\ee
where $I_F(p,m_q)$ is the scalar bubble integral with external
momentum $p$, but evaluated with the fermion mass and with 
Fermi-Dirac distribution, $f_F^\pm(\omega)=1/(\exp(\beta(\omega\mp\mu))+1)$:
\be
I_F(p,m_q)=I_F^0(p,m_q)-I_F^\beta(p,m_q),
\ee
\be
I_F^0(p,m_q)=\int\frac{d^4k}{(2\pi)^4}\frac{i}{(k^2-m_q^2+i\epsilon)
  ((k-p)^2-m_q^2+i\epsilon)},
\ee
\bea
\displaystyle
I_F^\beta(p,m_q)&=&\int\frac{d^3k}{(2\pi)^3}\frac{1}{4\omega_1\omega_2}
\left[\frac{f_F^+(\omega_1)+f_F^-(\omega_2)}
{p_0-\omega_1-\omega_2+i\epsilon}-\frac{f_F^-(\omega_1)+f_F^+(\omega_2)} 
{p_0+\omega_1+\omega_2+i\epsilon}\right.\nonumber\\
&&\qquad\qquad\qquad\quad
\left.-\frac{f_F^+(\omega_1)-f_F^+(\omega_2)}{p_0-\omega_1+\omega_2+i\epsilon}+
\frac{f_F^-(\omega_1)-f_F^-(\omega_2)}{p_0+\omega_1-\omega_2+i\epsilon}
\right].
\eea 

Substituting (\ref{Eq:bubble-tadpole}) into (\ref{Eq:pion}) and making use of 
(\ref{Eq:EoS}) one finds that for $p^2=0$ the inverse propagator vanishes. 
This means that Goldstone's theorem is fulfilled and a zero mass pion 
can be used consistently for the evaluation of the pion tadpole in 
(\ref{Eq:EoS}).

Next, we evaluate EoS, for instance, with dimensional regularisation. 
After $\overline{MS}$ subtraction, which corresponds to a perturbative 
renormalisation of $\lambda$ with a counterterm proportional to $g^4$, 
one arrives for EoS at
\be
m^2+\frac{\lambda}{6}\left(\Phi^2+T_{B}(0)\right)-\frac{4g^2N_c}
{\sqrt{N}}T_F(m_q)=0, 
\label{Eq:renEoS}
\ee 
where $T_a(x),~a=B,F$ stands for the bosonic/fermionic tadpole
contribution with mass $x$:
\bea
&
\displaystyle
T_a(x)=\frac{x^2}{16\pi^2}\ln\frac{ex^2}{M_{0a}^2}-T_a^\beta(x),\qquad
T_a^\beta(x)=\frac{1}{2\pi^2}\int_x^\infty
d\omega\sqrt{\omega^2-x^2}f_a(\omega),\nonumber\\
&
\displaystyle
f_F(\omega)=\frac{1}{2}\left[f_F^+(\omega)+f_F^-(\omega)\right],\quad
f_B(\omega)=\frac{1}{e^{\beta\omega}-1}.
\eea
Here $M_{0a}$ are the renormalisation scales, which can be chosen different 
for the fermionic and the bosonic tadpoles. Below we shall characterize 
the relationship of the two scales by $\eta=\ln(M_{0B}/M_{0F})$, 
which introduces an extra finite subtraction into the renormalisation scheme. 
We shall check if substantially different phenomenology can be achieved
when this freedom is exploited.

\section{Analytical determination of the TCP}
When looking for the location of continuous phase transitions in the
$T-\mu$ plane, one can study the left hand side of (\ref{Eq:renEoS}) 
for small field values and in power series of the fugacity, $\exp (\mu /T)$. 
One expands the distribution function for fermions and antifermions in power 
series, and changes the order of the integration and the summation, doing the 
integration first. One expands the resulting modified Bessel functions up to 
linear order in $m_{q}$ and eventually performs the remaining sum.
Remarkably, the logarithmic dependence on $\Phi$ is cancelled, and a 
power series in  $\Phi$ is the result. The sign for a continuous 
phase transition, the vanishing of the constant term is given 
(after substituting $N=4$) by the equation
\be
m^2+\frac{\lambda}{72}T_c^2
-\frac{g^2T_c^2}{2\pi^2}N_c\left(Li_2(-e^{\mu/T_c})+
Li_2(-e^{-\mu/T_c})\right)=0.
\label{Eq:const}
\ee
In particular for $\mu=0$  one finds a shift downwards in $T_c$ due to the
fermionic fluctuations:
\be
m^2+\left(\frac{\lambda}{72}+\frac{g^2N_c}{12}\right)T_c^2=0.
\ee
For small $\mu$ the coefficient of the quadratic part is positive,
therefore $T_c$ is consistently interpreted as a critical point. 
The second order line ends at $T=T_{TCP},~\mu=\mu_{TCP}$, where also 
the quadratic coefficient changes sign:
\be
\frac{\lambda}{6}+\frac{g^4N_c}{4\pi^2}\left[\frac{\partial}{\partial n}
\Big( Li_n(-e^{-\mu_{TCP}/T_{TCP}})+Li_n(-e^{-\mu_{TCP}/T_{TCP}})
\Big)\Big|_{n=0}- \ln\frac{cT_{TCP}}{M_0B}\right]=0,
\label{Eq:quadrat}
\ee
($\ln (c/2)=1-\gamma_E+\eta$).
The first order line for larger $\mu$ values was determined numerically 
directly from (\ref{Eq:renEoS}). Also the perfect agreement of the 
semi-analytically determined second order portion of the phase boundary 
with the exact one was verified numerically. 

The parameters $\lambda, ~g, ~\Phi, ~M_{0a}, ~m^2$ should be determined 
from the $T=0$ data. One requires the EoS to give $\Phi=f_\pi/2$, with
$f_\pi=93$ MeV. Then the value of $m_q(T=0)\equiv m_{q0}$ determines $g$. 
We have chosen $m_{q0}=m_N/3\approx312.67$, so that $g=6.72$.
The EoS at $T=0$ provides one relation among $m^2,~\lambda$ 
and the renormalisation scales $M_{0a}$, from which $m$ can
be determined once the other three parameters are known. 
One would like to fully fix these three parameters $M_{0a}$ and $\lambda$,
for instance, by studying the spectroscopy of the $\sigma$ particle 
\cite{szep02}. The relationship between the pole and the spectral function 
will be discussed in section 4. Here we elaborate on the features of the
phase boundary when scanning through a large domain of $M_{0B}$ and
$\eta$ for fixed value of $\lambda=400$.
 
\begin{figure}[htbp]
\begin{center}
\includegraphics*[width=7.75cm]{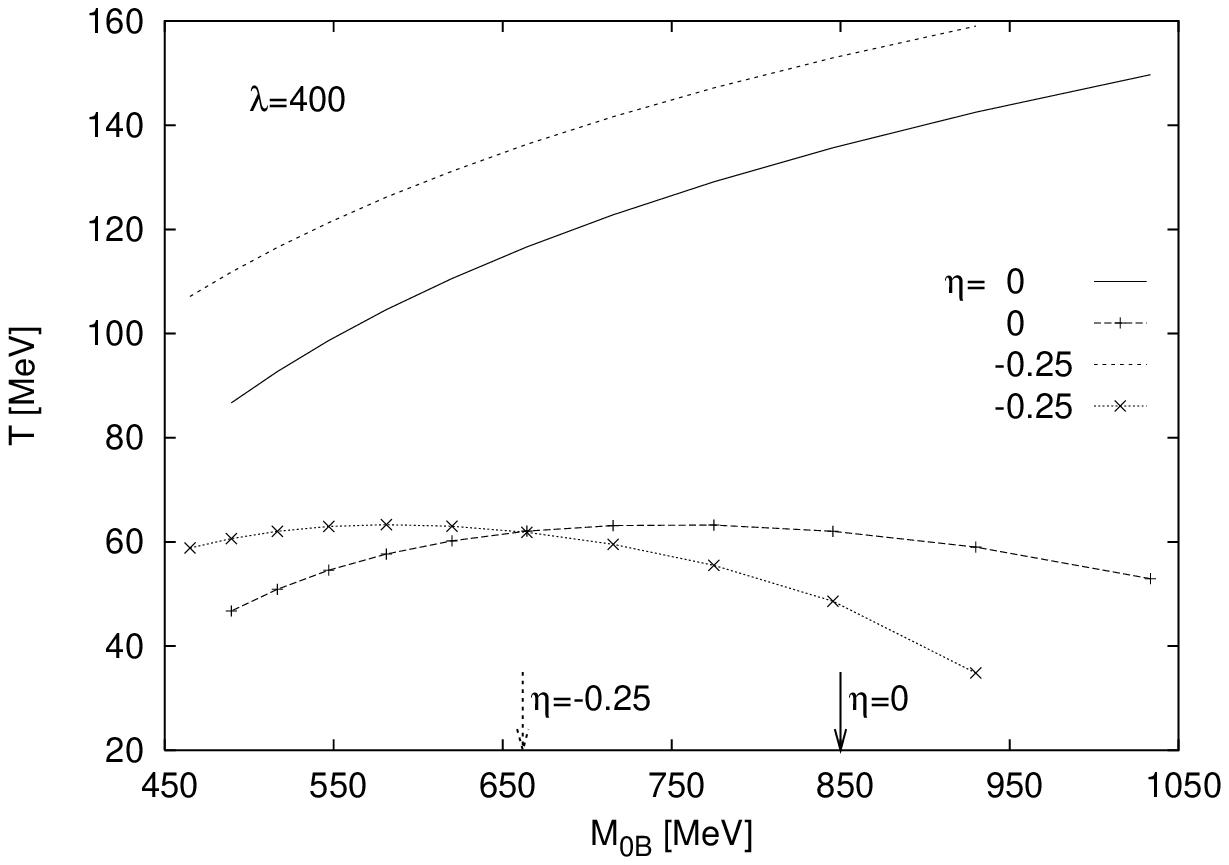}
\includegraphics*[width=7.75cm]{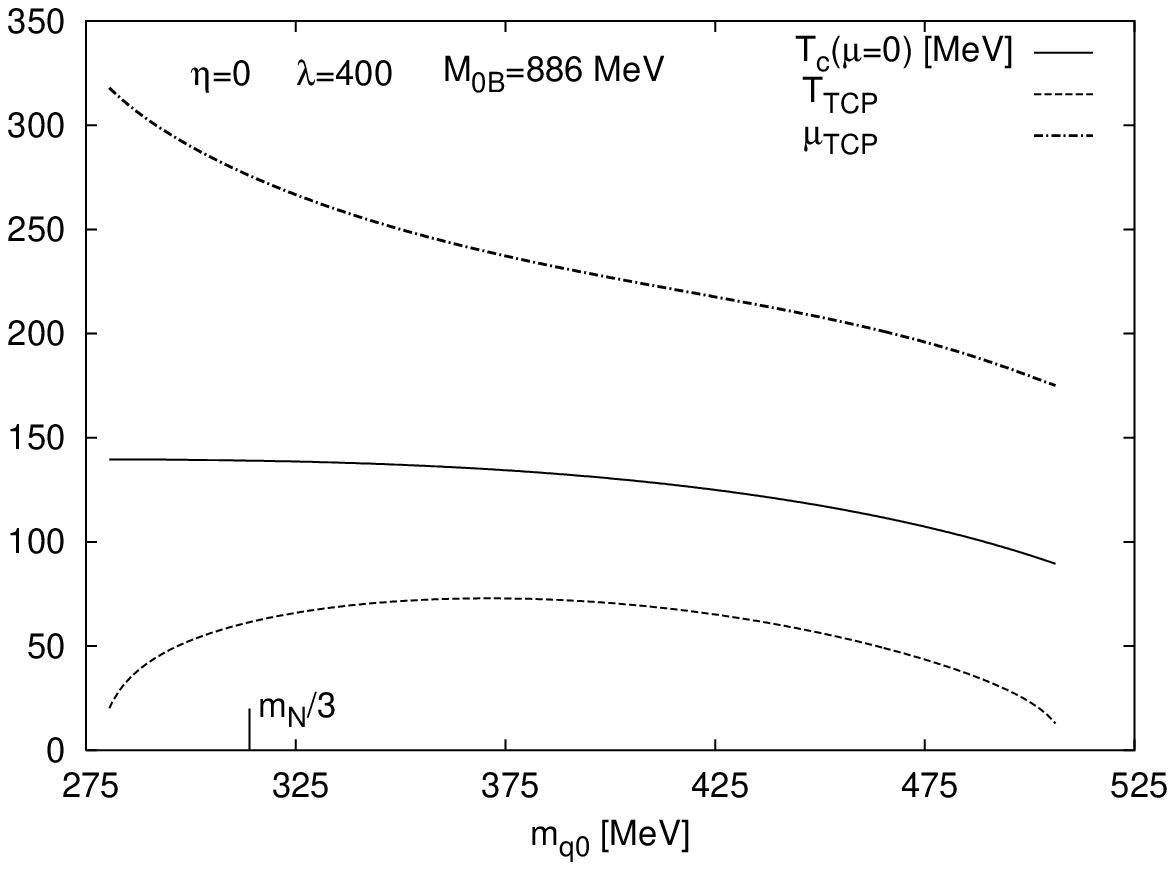}
\end{center}
\caption{L.h.s.: the variation of $T_c(\mu =0)$ (upper curves) and of $T_{TCP}$
(lower curves) with $M_{0B}$ for $\eta=0$ and $-0.25$ 
(for the definitions see the text). The arrows on the abscissa
separate the region of stable (right) and unstable (left) solutions
for different values of $\eta$. R.h.s.: the dependence of
$T_c(\mu=0), T_{TCP}$ and $\mu_{TCP}$ on the constituent quark mass $m_{q0}.$}
\label{Fig:C-TCP}
\end{figure}

In Fig.\ref{Fig:C-TCP} (l.h.s.) the dependence of the critical temperature on
$M_{0B}$ is shown at $\mu=0$ for $\eta=0,-0.25$. It should be
emphasized that by the arguments to be presented in Section 4 the allowed
range of $M_{0B}$ lies to the right from the arrow, appearing
on the horizontal axis of the figure. It is clear that in this region
one can arrange very easily $T_c(\mu=0)$ to occur in the
region 150-170 MeV, where it is indicated by lattice studies of
the two flavor QCD with dynamical quarks \cite{karsch01,alikhan01}. 
The temperature of the tricritical point changes very mildly and stays 
for all choices below 70 MeV. Another informative representation of the 
phase structure is to display the dependence of 
$T_c(\mu=0), T_{TCP}$ and $\mu_{TCP}$ on the constituent quark
mass, which appears on the r.h.s. of Fig.~\ref{Fig:C-TCP}.

\begin{figure}[htbp]
\begin{center}
\includegraphics*[width=9cm]{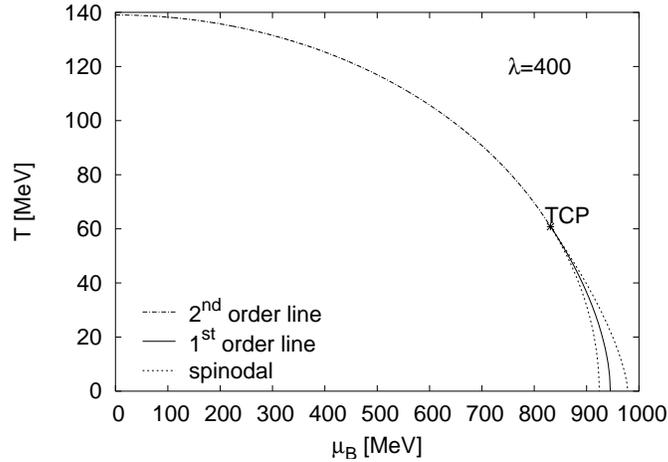}
\end{center}
\caption{The $T-\mu$ phase diagram in the chiral limit for
    $M_{0B}=886$ MeV and $\eta=0$.}
\label{Fig:TCP}
\end{figure}

We draw the complete phase diagram in Fig.~\ref{Fig:TCP} 
with $m_{q0}=m_{N}/3$ and $\eta=0$ for $M_{0B}=886$ MeV. The left
spinodal is simply the continuation of the second order phase
boundary, the right spinodal is found by locating the turning point of
the two-valued order parameter curve as a function of $\mu$ for fixed $T$.
We notice that the metastable region between the spinodal curves
becomes narrower with increasing $M_{0B}$. In between we draw the phase
boundary by finding the degeneracy point of the thermodynamical potential.
When compared with  the features of the previous (non-renormalised) 
investigations in effective models \cite{scavenius01}, one realizes that the 
theoretical consistency achieved in the present treatment does not produce 
qualitative differences. If we tune $\eta$ in such a way that $T_c(\mu=0)$
increases then TCP moves further down and rightwards on the
  phase diagram. Our phase boundary drops much steeper than
the present lattice estimates \cite{katz03} performed with rather
heavy quarks and is quite close to the freeze-out curve of 
Cleymans and Redlich \cite{cleymans99}.

\section{Poles in the $\sigma$ channel and their relation to the 
spectral function }

In the present letter we need the spectral function in the 
$\sigma$ channel only for $T=0$, that is for discussing the
sensitivity of its pole structure 
to the selection of the couplings, especially the renormalisation scales
of the effective model. A detailed analysis of the features of the
finite temperature pole trajectory and the phenomenon of the 
threshold enhancement is postponed to a forthcoming publication.
The large $N$ form of the inverse propagator of the $\sigma$ particle
\cite{szep02} is now modified by the contribution of the fermionic loop:
\bea
iG^{-1}_\sigma(p)&=&p^2-m^2-\frac{\lambda}{2}\Phi^2-\frac{\lambda}{6N}
\lag\pi^a\pi^a\rag_M-S_{chain}(p)\nonumber\\
&-&\frac{gN_c}{N\Phi}\lag\overline{\psi}\psi\rag+
\frac{2g^2N_c}{\sqrt{N}}(4m_{q0}^2-p^2)I_F(p,m_{q0})+
{\rm counterterms}.
\label{Eq:sigma_prop}
\eea
Here $S_{chain}(p)$ represents the sum of the infinite sequence of
pion bubble contributions to the $\sigma$ self energy.
This equation simplifies if (\ref{Eq:EoS}) is taken into account to
\be
iG^{-1}_\sigma(p)=p^2-\frac{\lambda}{3}\Phi^2-S_{chain}(p)+
\frac{2g^2N_c}{\sqrt{N}}(4m_{q0}^2-p^2)I_F(p,m_{q0})+{\rm counterterms}.
\ee

Fermionic counterterms are needed for the perturbative renormalisation
of the coefficients of $p^2$ and of $\Phi^2$, both receiving divergent
contribution from $I_F(p,m_{q0})$. Using $\overline{MS}$ subtraction the 
fermionic part of the coupling constant counterterm which is 
${\mathcal O}(g^4)$, proves to be the same as the one encountered
in the renormalisation of the EoS. Additional bosonic
counterterms form a power series in $\lambda$, and are needed 
to renormalise the chain of bubbles $S_{chain}$. This series can be formally
resummed, resulting in a non-perturbative renormalisation of the meson
sector \cite{szep02}.
The relation between the bare and renormalised coupling constant is:
\be
\displaystyle
\lambda_{\rm bare}=\frac{\lambda}{1+\frac{\lambda}{6}b_{div}}+
\frac{24 g^4}{\sqrt{N}} b_{div},
\qquad
b_{div}=\frac{1}{16\pi^2}\left(-\frac{1}{\epsilon}+\gamma-2-\ln (4\pi)\right).
\ee

Summing up the renormalised terms of the bubble chain, the final renormalised 
$\sigma$ propagator at rest (${\bf p}=0$) reads as follows:
\be
iG^{-1}_\sigma(p_0,0)=p_0^2
-\frac{\lambda}{3}\Phi^2\left(1-\frac{\lambda}{6}I_{B,R}(p_0,0)\right)^{-1}
+\frac{2g^2N_c}{\sqrt{N}}(4m_{q0}^2-p_0^2)I_{F,R}(p_0,m_{q0}).
\label{Eq:pole}
\ee
The spectral function is obtained by putting 
$p_0\rightarrow p_0+i\epsilon$, and taking the imaginary part along the
physical real axis:
\be
\rho_\sigma(p_0)=\frac{1}{\pi}{\rm Im}\left[iG_\sigma 
(p_0+i\epsilon,0)\right].
\ee

Below we explore how the $\sigma$ pole structure in the lower 
halfplane is reflected in the shape of the spectral function
for a broad range of the values 
of the renormalisation scale $M_{0B}$ as it appears in
Fig.~\ref{Fig:C-TCP}. An analytic continuation of (\ref{Eq:pole}) 
is understood via the $T=0$ expressions of $I_{a,R}$, 
which for complex values of $p_0$ can be written explicitly as
\bea
I_{B,R}^0(p_0,0)&=&
\frac{1}{16\pi^2}\left[\ln\frac{p_0^2}{M_{0B}^2}-ik_1\pi\right]
\nonumber\\
I_{F,R}^0(p_0,m_{q0})&=&
\frac{1}{16\pi^2}
\left[\ln\frac{m_{q0}^2}{M_{0F}^2}-Q
\left(\ln\frac{1-Q}{1+Q}+ik_2\pi\right)\right],
\quad Q=\sqrt{1-\frac{4m_{q0}^2}{p_0^2}}.
\label{Eq:bubbles}
\eea
The above formulae correspond to analytical continuations onto 
specific Riemann-sheets characterised by two integers $(k_1,k_2)$. 
For example $k_1=1, k_2=1$ and $k_1=1, k_2=-1$ corresponds to
an analytical continuation across the real axis above and below the
two-fermion threshold, respectively.
We found roots of the right hand side of (\ref{Eq:pole}) for all
$k_1>0,~k_2<0$ integers. For a wide range of $\lambda$ values
the real parts of all poles have nearly the same value and lie slightly
below the two-quark threshold. Therefore one expects the spectral
function to have the maximum of its ``resonant'' part at
$p_{0,max}\approx {\rm Re}(p_{0,pole})$.

This expectation turned out to be incorrect below certain
values of the renormalisation scale, depending on $\eta$.
What spoils this feature is the Landau pole structure on 
the first Riemann sheet, which is an indication of the effective nature of the 
theory. This structure sensitively depends on $M_{0B}$. 
Apart from the well known large scale Landau pole(s) on the imaginary axis, 
which decreases exponentially with increasing values of $\lambda$, new 
small scale poles might emerge.

The mechanism for  their occurrence is the cleanest to illustrate
for $\lambda=0$. In this case the imaginary part of a physical
$\sigma$ pole necessarily vanishes. For $M_{0B}> M_{0L}$ the pole is
real but its location shifts towards the origin as $M_{0B}$ decreases.
At $M_{0B}=M_{0L}$ 
it reaches the origin and collides with its mirror
arriving along the real axis from the direction $p_0<0$. 
For $M_{0B}<M_{0L}$ a conjugate pair of poles appears
along the imaginary axis. One of them signals the instability of the
system at very low energy scales, completely invalidating our
solution in this region of the renormalisation scales. This dipole
induced instability prevails also for $\lambda =400$, restricting
the choice of the renormalisation scale to $M_{0B}>M_{0L}$. 
One finds from (\ref{Eq:bubbles}) $M_{0L}=m_{q0}\exp(\eta+1)$ which,
with our present numerical values of $f_\pi$ and $m_{q0}$, gives
$M_{0L}=849.92$ MeV at $\eta=0$ and
$M_{0L}=661.92$ MeV at $\eta=-0.25$. In the allowed region $M_{0B}<M_{0L}$,
$|M_{Landau}|/|p_{0,pole}|$ stays always well above unity. There, the
real part of the pole position and the maximum of the spectral
function coincide as expected.

\begin{figure}[t]
\begin{center}
\includegraphics*[width=9cm]{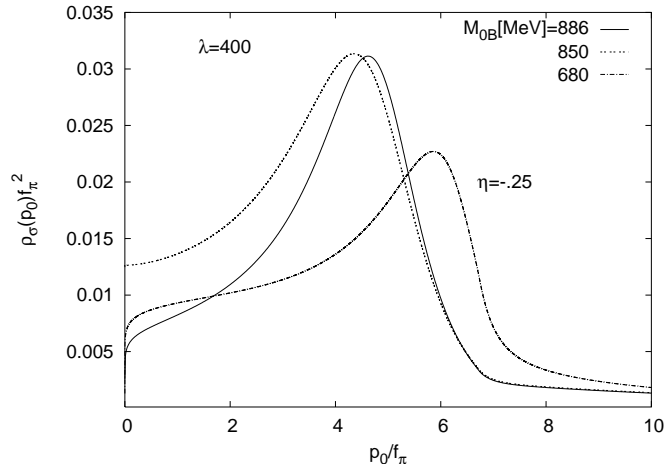}
\end{center}
\caption{The $T=0$ spectral function for different allowed
choices of $M_{0B}$ and $\eta$ in the chiral limit. 
The variation is restricted to the stability region of the system.}
\label{Fig:spectral}
\end{figure}

In Fig. \ref{Fig:spectral} we display the emerging spectral functions for
$M_{0B}=850, 886$ MeV and $\eta=0$. In addition also the curve for
$M_{0B}=680$ MeV and $\eta=-0.25$ is shown. The effect of the fermions 
brings the location of the ``$\sigma$-peak'' closer to the
phenomenologically expected region: at $\eta=0$ we find 
$p_{0,peak}\leq 430$ MeV, while the corresponding poles are
located at ${\rm Re}p_{0,pole}\simeq 450$ MeV. 
The $\Gamma_\sigma/M_\sigma$ ratios still turn out to be
lower than expected by 40-50\%.
Finally, one might remark that the stability of the broken symmetry
phase also imposes a lower limit on $M_{0B}$, which is, however, lower
in our case than the bound, described above.

Two short notes are of order in connection with the thermal evolution
of the poles. First, near $T_c$ all ``$\sigma$-poles'' flow to the origin
of the $p_0$ plane asymptotically along the same curve. The real part
of this trajectory scales with $\Phi (T)$, the imaginary part with
$\Phi^2(T)$. This means that the inclusion of the fermions qualitatively 
changes the dynamics of the system, since in the pure $O(N)$ model
the trajectory reached the origin along the imaginary axis. 
Second, for the scaling exponent of the order parameter,
$\Phi (T)\sim (T-T_c)^\beta$, on the full second order line we find
its mean field estimate 1/2. For the single point of TCP $\beta =1/4$ was
found within the numerical accuracy of its determination, in full
agreement with the standard result of statistical physics
\cite{gebhard}. At TCP only logarithmic corrections are expected to
this value.

\section{Conclusions and discussion}

In this letter we have presented a theoretically fully consistent
derivation of the phase structure in the $T-\mu$ plane of strong matter
from the chiral $\sigma$ model in the chiral limit using an
approximate solution valid to ${\mathcal O}(1/\sqrt{N})$ when the number
of the Goldstone bosons ($N$) goes to infinity. 
A novel feature arising from including the fermions is the strong
influence of the renormalisation scheme on the shape of the spectral function.
It is essential to check also the dependence of the excitation spectra
  on the renormalisation scale, which allows to restrict considerably
  the range for the acceptable values of $M_{0B}$.
The predictions on the phase boundary in the $T-\mu$
plane depend on the renormalisation scale(s) only smoothly, leading to a unique
semi-quantitative characterisation of the phase structure. 
In the allowed range of the parameters one finds $T_c(\mu=0)$ in
reasonable agreement with the results of MC simulations, while one encounters
a genuinely low value for $T_{TCP}\approx 60-70~ {\rm MeV}$.

Our approach can be further improved by the dynamical determination of the 
fermion mass from a one-loop Schwinger-Dyson equation. For a realistic 
comparison with hot heavy-ion systems it is important also to include the 
effect of explicit chiral symmetry breaking. Possible dependence on the 
renormalisation scheme can be optimised by requiring the best phenomenological 
description of the scalar-isoscalar spectral function for $T=0$.
These extensions require a careful implementation of the proposed 
(non-perturbative) renormalisation procedure, which will prove more
complex than in the present case. Our final goal is the application of the 
large $N$ method to the three-flavor $SU(3)_L\times SU(3)_R$ 
case with realistic constituent quark masses.
 
\begin{ack}
The authors acknowledge the support of the Hungarian Research Fund (OTKA)
under contract numbers F043465, T034980, and T037689.
\end{ack}


\begin{thebibliography}{9}
\bibitem{fodor02a}Z. Fodor and S.D. Katz, Phys. Lett. {\bf B534} (2002) 87
\bibitem{fodor02}Z. Fodor and S.D. Katz, JHEP 0203 (2002) 014
\bibitem{karsch02}C.R. Allton {\it et al.}, Phys. Rev. {\bf D66}
  (2002) 074507
\bibitem{philipsen02} P. de Forcrand and O. Philipsen,
  Nucl. Phys. {\bf B642} (2002) 290
\bibitem{lombardo03} M. d'Elia and M.-P. Lombardo, Phys. Rev. {\bf
  D67} (2003) 014505 
\bibitem{katz03}S.D. Katz, {\it Lattice QCD at finite $T$ and $\mu$}, 
  hep-lat/0310005
\bibitem{wiese03}S. Chandrasekaran, M. Pepe, T.D. Steffen and
  U.-J.Wiese, {\it Non-linear Realisation of Chiral Symmetry on the
  Lattice}, hep-lat/0306020
\bibitem{redlich03a}F. Karsch, K. Redlich and A. Tawfik,
  Eur. J. Phys. {\bf C29} (2003) 549
\bibitem{redlich03b}F. Karsch, K. Redlich and A. Tawfik,
  Phys. Lett. {\bf B571} (2003) 67
\bibitem{dumitru03}A. Dumitru, D. R{\"o}der and J. Ruppert, hep-ph/0311119
\bibitem{barducci94}A. Barducci, R. Casalbuoni, S. DeCurtis, R.Gatto
  and G. Pettini, Phys. Rev. {\bf D49} (1994) 426
\bibitem{berges99}J. Berges and K. Rajagopal, Nucl. Phys. {\bf B538}
  (1999) 215
\bibitem{scavenius01}O. Scavenius, \'A. M{\'o}csy, I.N. Mishustin and
  D.H. Rischke, Phys. Rev. {\bf C64} (2001) 045202
\bibitem{fujii03}H. Fujii, Phys. Rev. {\bf D67} (2003) 094018
\bibitem{hatsuda99}T. Hatsuda, T. Kunihiro and H. Shimizu,
  Phys. Rev. Lett. {\bf 82} (1999) 2840
\bibitem{lenaghan00}J.I. Lenaghan and D.H. Rischke, J. Phys. {\bf G26}
  (2000) 431; D. R{\"o}der, J. Ruppert and D.H. Rischke, Phys. Rev. {\bf
  D68} (2003) 016003
\bibitem{collins87}J.C. Collins, {\it Renormalisation}, Cambridge
  University Press, 1987
\bibitem{blaizot03}J.-P. Blaizot, E. Iancu and U. Reinosa,
  Phys. Lett. {\bf B568} (2003) 160
\bibitem{vanhees02}H. van Hees and J. Knoll, Phys. Rev. {\bf D66}
  (2002) 025028
\bibitem{szep02}A. Patk{\'o}s, Zs. Sz{\'e}p and P. Sz{\'e}pfalusy,
  Phys. Lett. {\bf B537} (2002) 77
\bibitem{szepfalusy02}A. Patk{\'o}s, Zs. Sz{\'e}p and P. Sz{\'e}pfalusy,
  Phys. Rev. {\bf D66} (2002) 116004
\bibitem{dolan74}L. Dolan and R. Jackiw, Phys. Rev. {\bf D9} (1974) 3320
\bibitem{karsch01}F. Karsch, E. Laermann and A. Peikeert,
  Nucl. Phys. {\bf B605} (2001) 579
\bibitem{alikhan01}A. Ali Khan {\it et al.} [CP-PACS Collaboration]
  Phys. Rev. {\bf D63} (2001) 034502
\bibitem{cleymans99}J. Cleymans and K. Redlich, Phys. Rev. {\bf C60}
  (1999) 054908
\bibitem{gebhard} W. Gebhardt and U. Krey, Phasen\"uberg{\"a}nge und
 Kritische Phenom{\"a}ne, F. Vieweg \& Sohn, Braunschweig, 1980
\end{thebibliography}
\end{document}